# Converging an Overlay Network to a Gradient Topology

Håkan Terelius*, Guodong Shi*, Jim Dowling†, Amir Payberah†, Ather Gattami* and Karl Henrik Johansson*

*KTH - Royal Institute of Technology, {hakante,guodongs,gattami,kallej}@kth.se

†Swedish Institute of Computer Science (SICS), {jdowling,amir}@sics.se

*Abstract*— In this paper, we investigate the topology convergence problem for the gossip-based Gradient overlay network. In an overlay network where each node has a local utility value, a Gradient overlay network is characterized by the properties that each node has a set of neighbors with the same utility value (a similar view) and a set of neighbors containing higher utility values (gradient neighbor set), such that paths of increasing utilities emerge in the network topology. The Gradient overlay network is built using gossiping and a preference function that samples from nodes using a uniform random peer sampling service. We analyze it using tools from matrix analysis, and we prove both the necessary and sufficient conditions for convergence to a complete gradient structure, as well as estimating the convergence time and providing bounds on worst-case convergence time. Finally, we show in simulations the potential of the Gradient overlay, by building a more efficient live-streaming peer-to-peer (P2P) system than one built using uniform random peer sampling.

## I. INTRODUCTION

Recent years have witnessed growing interest in using randomized gossiping algorithms to build distributed systems, in particular in the areas of overlay networks, sensor networks and cloud computing storage services [1], [2]. Gossip-based, or pair-wise exchange, algorithms have primarily been used to implement aggregation algorithms, information dissemination, peer sampling (the uniform random sampling of a node from the set of all nodes in a P2P system), and to construct overlay network topologies. Much of the existing analysis of gossip-based algorithms has focused on the convergence properties of aggregation algorithms and peer sampling services, for both fixed topologies [3] and regular graphs [4], [5].

However, research in gossiping has also focused on using the Preferential Connectivity Model [6] to construct overlay network topologies, where nodes connected initially in a random graph use a preferential connection function to break the symmetry of the random graph and build a topology that contains useful global information. Barabasi first described how a preferential attachment function in a growing network can build a scale-free network topology from a random graph [7]. In particular, he showed how the power-law distribution of links in the the World Wide Web can emerge when arriving nodes preferentially attach to existing nodes with higher edge degree. Information about the structure of the Web's topology is currently used, among other things, to build more efficient search algorithms. Barabasi's preferential attachment functions are based on global state (the in-degree of nodes). However, in overlay networks, nodes have only a relatively small partial view of the system, so preference functions are based only on local state and the state of the node's neighbors. Examples of existing overlay networks that construct their topologies using gossiping and preference functions include Spotify, that preferentially connects nodes with similar music play-lists [8], Sepidar, that preferentially connects P2P live-streaming nodes with similar upload bandwidth capacity [9], and T-Man, a framework that provides a generic preference function for building such overlays [10].

To the best of our knowledge, there has been no analysis of the convergence properties of such information-carrying gossip-generated topologies built using preference functions. These systems, however, do not require the growth of a network to construct a new topology, as systems are constantly updated using a peer sampling service. In this paper, we introduce an analysis of the convergence properties for the Gradient overlay network. The Gradient topology belongs to this class of gossip-generated overlay networks that are built from a random overlay by symmetry breaking using a preference function. Formally, a Gradient topology is defined as an overlay network where, for any two nodes $p$ and $q$ that have local utility values $U(p)$ and $U(q)$, if $U(p) \geq U(q)$ then $dist(p, r) \leq dist(q, r)$, where $r$ is a (or the) node with highest utility in the system and $dist(x, y)$ is the shortest path length between nodes $x$ and $y$ [11]. In the Gradient overlay, nodes have two preference functions that build two sets of neighbors: a similar view and a gradient view. For the similar view, nodes prefer neighbors with closer utility values, while for the gradient view, nodes prefer nodes with higher, but closer, utility values. Together these preference functions build a topology where gradient paths of increasing utilities emerge in the system [12], see figure 1.

Our analysis of the Gradient overlay, involves proving that the preference functions cause the system topology to converge to a gradient structure. We also establish bounds on the worst-case convergence rate for a given initial graph. Finally, we show in simulations how the Gradient structure is used to build a more efficient live-streaming system than one built using uniform random peer sampling.

## II. PROBLEM SETUP

Consider a network whose topology can be described by a directed graph $\mathcal{G}(\mathcal{N}, \mathcal{E})$. Each node in the network is represented by a vertex in the graph, and each link is represented by a directed edge (see figure 1(a)). We denote the vertex set by $\mathcal{N} = \{1, \ldots, N\}$, where each node $i$ is

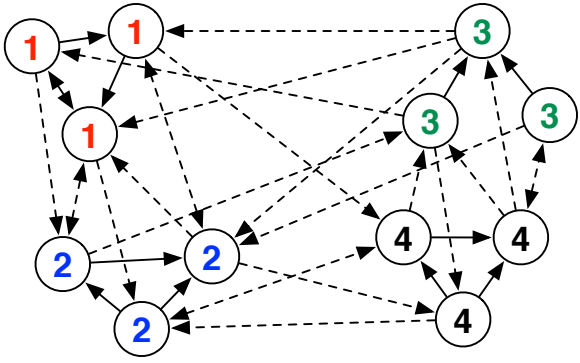 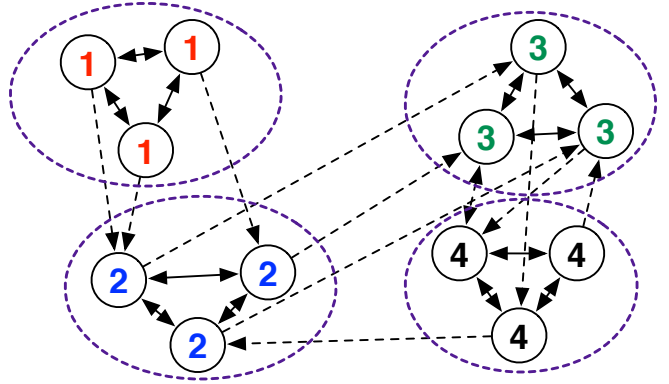

(a) The initial graph.   (b) The graph after converging to a gradient topology.

Fig. 1. The network is described as a directed graph. The nodes are labeled with their respective utility value, and the edges from the similar neighbor set are shown. Solid edges are used between nodes with equal utility value, and dashed edges between nodes with different utility value.

given a utility value $U(i) \in \Lambda$ from a given utility value set $\Lambda = \{1, \ldots, n\}$.

Let $\Lambda_u \triangleq \{i \mid U(i) = u\}$ be the set consisting of nodes with utility $u$, $u = 1, \ldots, n$. Suppose $|\Lambda_u| = m_u$, where $|\cdot|$ represent the number of elements for a finite set. The utility distance function is denoted as $d(i,j) \triangleq |U(i) - U(j)|$.

The neighbor set $N_i(t)$ of node $i$ at time $t$ consists of two parts, the similar view $N_i^s(t)$ and the random view $N_i^r(t)$. Nodes in the similar view are supposed to be the neighbors whose utility values are close to $U(i)$, while nodes in the random view are a random sample of the nodes in the network.

**Assumption 1:** For every node $i \in \mathcal{N}$, if $i \in \Lambda_u$ then $i$ has exactly $m_u$ similar neighbors, excluding itself.

$$N_i^s(t) = \{i_1, i_2, \cdots, i_{m_u}\}$$

## III. TOPOLOGY DYNAMICS

For any given initial graph, consider the following algorithm for the topology dynamics:

**Algorithm 1.** Let $t = 1$.

*Step 1.* At time $t$, node $i$ chooses a random neighbor $j$ from node set $\mathcal{N}$ with equal probability, i.e.,

$$\mathcal{P}\{N_i^r(t) = \{j\}, j \in \mathcal{N}\} = p,$$

where $p$ satisfies $0 < Np < 1$. Notice that the random neighbor set is empty with probability $1 - Np$, in which case we skip Step 2.

*Step 2.* If the random node $j$ is an improvement of the similar neighbor set, then we replace the worst node in $N_i^s$ with $j$. Thus, if $U(j) \geq U(i)$ and $d(i,j) \leq \max_{k \in N_i} d(i,k)$, then add $j$ to $N_i^s$ and remove $u = \arg\max_{k \in N_i^s} d(i,k)$ from $N_i^s$.

*Step 3.* Let $t = t + 1$, then go to Step 1.

This paper considers the problem of whether the system topology will converge to a gradient structure with the proposed algorithm, and the convergence rate for a given initial graph.

For every node $i \in \mathcal{N}$, we define

$$X_t^{(i)} \triangleq \sum_{j \in \mathcal{N}_i^s(t)} \mathrm{sgn}(d(i,j)),$$

where

$$\mathrm{sgn}(v) = \begin{cases} 0, & \text{if } v = 0 \\ 1, & \text{otherwise} \end{cases}.$$

Thus, $X_t^{(i)}$ counts the number of nodes in $i$'s similar neighbor set with a different utility value than $U(i)$.

Let $\mathcal{G}(t)$ be the graphs generated by Algorithm 1. Then we give the definition of gradient convergence as follows (see also figure 1).

*Definition 3.1:* $\mathcal{G}(t)$ is said to converge to a gradient topology if $\lim_{t \to \infty} X_t^{(i)} = 1$ for $i \in \mathcal{N}$, and with $U(j) = U(i) + 1$, where $j$ is the only node with different utility in $\lim_{t \to \infty} N_i^s(t)$, for $i \in \mathcal{N}$, $U(i) < n$.

## IV. CONVERGENCE ANALYSIS

In this section, we propose a gradient convergence analysis, where we focus on the first condition $\lim_{t \to \infty} X_t^{(i)} = 1$. The analysis can be extended to handle the second condition, with similar results. Since each node updates its neighbor set independently, the analysis on $X_t^{(i)}$ can be carried out respectively. Therefore, we let $X_t$ represents $X_t^{(i)}$, $i \in \mathcal{N}$, in the following discussions to simplify the notations.

Denote $m = \max_u\{m_u\}$. Then it is not hard to see that $X_0 = m$ is the worst initial condition. In practice, the sampling probability $p$ in Algorithm 1 can be time-varying, i.e., $p = p_t, t = 1, 2, \ldots$. Furthermore, for all $t = 1, 2, \ldots$, one has

$$\mathcal{P}\{X_{t+1} = k \mid X_t = k+1\} = kp_t \qquad (1)$$

where $kp_t$ is the probability of sampling one of the $k$ remaining nodes with the same utility value.

## A. Almost Sure Convergence

We propose a both necessary and sufficient condition on the probabilities $p_t$ for the convergence of Algorithm 1.

*Theorem 4.1:* The graph generated by Algorithm 1 converges to a gradient topology ($X_t = 1$) with probability 1 if and only if

$$\lim_{T\to\infty} \prod_{t=0}^{T}(1-p_t) = 0. \quad (2)$$

Before proving Theorem 4.1, let us take a closer look at Algorithm 1, and notice especially that the stochastic process (1) for $X_t$ has the Markov property, hence we can describe it as a Markov chain.

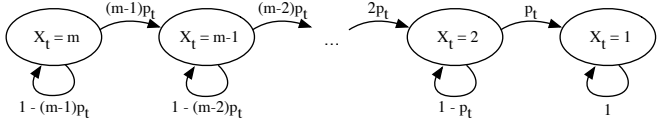

Let $\pi(t)$ denote the (row vector) probability distribution for the states $X_t$, i.e.,

$$\pi_i(t) = \mathcal{P}\{X_t = i\}. \quad (3)$$

The evolution of $\pi(t)$ can be written in matrix form as

$$\pi(t+1) = \pi(t)P_t, \quad (4)$$

where $P_t$ is the transition matrix at time $t$,

$$P_t = \begin{bmatrix} 1-(m-1)p_t & (m-1)p_t & 0 & \cdots & 0 & 0 \\ 0 & 1-(m-2)p_t & (m-2)p_t & \cdots & 0 & 0 \\ 0 & 0 & 1-(m-3)p_t & \cdots & 0 & 0 \\ \vdots & \vdots & \vdots & \ddots & \ddots & 0 \\ 0 & 0 & 0 & \cdots & 1-p_t & p_t \\ 0 & 0 & 0 & \cdots & 0 & 1 \end{bmatrix}.$$

Since $P_t$ is a triangular matrix, the eigenvalues are given by the diagonal elements, i.e., the eigenvalues of $P_t$ are $\lambda_i(t) = 1-(m-i)p_t$, $i=1,\ldots,m$. Notice that $\lambda_m(t) = 1$, and all other eigenvalues are strictly less than one. Furthermore, all eigenvalues are distinct, hence the eigenvectors form a basis for $\mathbb{R}^m$. In the following lemma, we characterize the eigenvectors.

*Lemma 4.1:* The eigenvector $\xi^i(t)$ corresponding to eigenvalue $\lambda_i(t)$ is independent of $p_t \neq 0$, $i=1,\ldots,m$.

*Proof:* The (left-)eigenvectors of $P_t$ satisfy $\lambda_i(t)\xi^i(t) = \xi^i(t)P_t$. Let $\xi^i_j(t)$ denote the $j$:th component of $\xi^i(t)$, then

$$\begin{cases} (1-(m-i)p_t)\xi^i_1(t) = (1-(m-1)p_t)\xi^i_1(t) \\ (1-(m-i)p_t)\xi^i_j(t) = (1-(m-j)p_t)\xi^i_j(t) + \\ \quad (m-j+1)p_t\xi^i_{j-1}(t) \quad j=2,\ldots,m \end{cases}$$

$$\Downarrow$$

$$\begin{cases} (i-1)\xi^i_1(t) = 0 \\ (i-j)\xi^i_j(t) = (m-j+1)\xi^i_{j-1}(t) \quad j=2,\ldots,m \end{cases}$$

$$\Downarrow$$

$$\begin{cases} \xi^i_j(t) = 0 & \text{if } j < i \\ \xi^i_j(t)\frac{i-j}{m-j+1} = \xi^i_{j-1}(t) & \text{if } j > i \end{cases} \quad (5)$$

while $\xi^i_i(t)$ can be chosen as an arbitrary non-zero value. ∎

Lemma 4.1 implies especially that all $P_t$ are simultaneously diagonalizable, hence we can drop the parameter $t$ from $\xi^i$.

Let us now return to the initial probability distribution $\pi(0)$, and let us express it in the eigenvector basis as

$$\pi(0) = \sum_{i=1}^{m} \alpha_i \xi^i, \quad (6)$$

for some real numbers $\alpha_i$.

*Lemma 4.2:* $\alpha_m \xi^m = \mathbf{e}_m$, where $\mathbf{e}_i$ is the Cartesian unit vector $[0,\ldots,0,1,0,\ldots,0]^T$ with 1 in position $i$.

*Proof:* Let us consider $\xi^i \mathbf{1}$ for $i = 1,\ldots,m-1$. By equation (5),

$$\xi^i \mathbf{1} = \sum_{j=1}^{m} \xi^i_j = \sum_{j=i}^{m} \xi^i_j = \sum_{j=0}^{m-i} \xi^i_{i+j}$$

We will show by induction that

$$\sum_{j=0}^{k} \xi^i_{i+j} = \frac{m-i-k}{m-i}\xi^i_{i+k}. \quad (7)$$

The case when $k=0$ is clearly true, thus, assume (7) holds for $k$ and consider $k+1$,

$$\sum_{j=0}^{k+1} \xi^i_{i+j} = \sum_{j=0}^{k} \xi^i_{i+j} + \xi^i_{i+k+1} = \frac{m-i-k}{m-i}\xi^i_{i+k} + \xi^i_{i+k+1}$$
$$= \frac{m-i-k}{m-i}\frac{-(k+1)}{m-i-k}\xi^i_{i+k+1} + \xi^i_{i+k+1}$$
$$= \frac{m-i-(k+1)}{m-i}\xi^i_{i+k+1}$$

Using (7) implies that $\xi^i \mathbf{1} = 0$, $i=1,\ldots,m-1$, and further, $\pi(0)\mathbf{1} = \alpha_m \xi^m \mathbf{1}$. Since $\pi(0)$ is a probability distribution, we know that $\pi(0)\mathbf{1} = 1$, but (5) tells us that only the last component of $\xi^m$ is non-zero, hence the lemma follows. ∎

We are now ready to prove the main theorem.

*Proof:* (Theorem 4.1) The convergence condition is equivalent to $\lim_{T\to\infty} \pi(T) = \mathbf{e}_m$. Using (4) and (6) gives us

$$\pi(T) = \pi(0) \prod_{t=0}^{T-1} P_t = \sum_{i=1}^{m} \alpha_i \xi^i \prod_{t=0}^{T-1} P_t =$$
$$\sum_{i=1}^{m} \alpha_i \xi^i \prod_{t=0}^{T-1} \lambda_i(t) = \sum_{i=1}^{m-1} \alpha_i \xi^i \prod_{t=0}^{T-1} \lambda_i(t) + \mathbf{e}_m \quad (8)$$

Consider the limit $\lim_{T\to\infty} \pi(T)$,

$$\lim_{T\to\infty} |\pi(T) - \mathbf{e}_m| = \lim_{T\to\infty} \left|\sum_{i=1}^{m-1} \alpha_i \xi^i \prod_{t=0}^{T-1} \lambda_i(t)\right| \leq$$
$$\sum_{i=1}^{m-1} |\alpha_i \xi^i| \cdot \lim_{T\to\infty} \prod_{t=0}^{T-1}(1-p_t).$$

Clearly, this converges to zero if $\lim_{T\to\infty} \prod_{t=0}^{T}(1-p_t) = 0$.

Also, the set of initial probability distributions spawns $\mathbb{R}^m$, thus, there exists an initial probability distribution $\pi(0)$ such

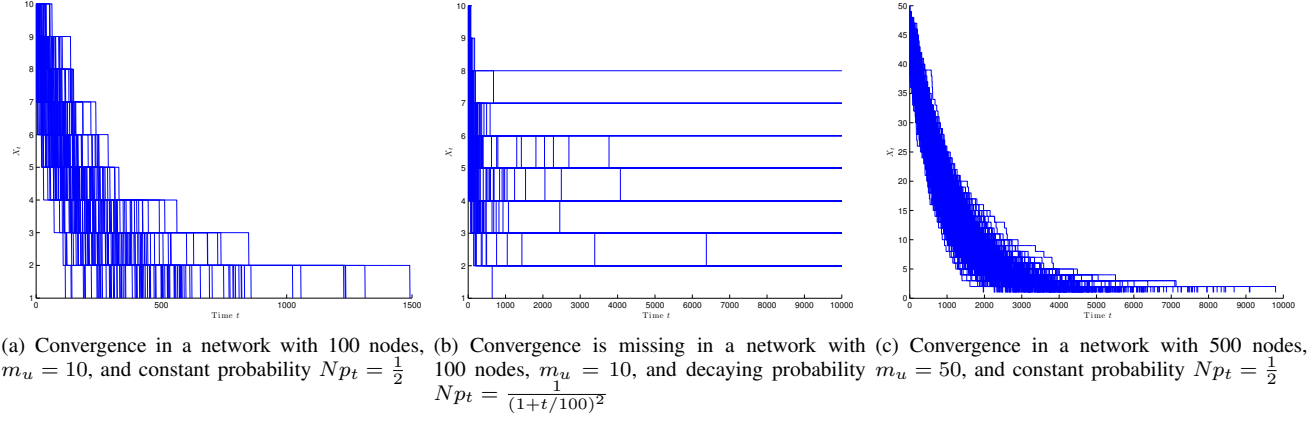

(a) Convergence in a network with 100 nodes, $m_u = 10$, and constant probability $Np_t = \frac{1}{2}$

(b) Convergence is missing in a network with 100 nodes, $m_u = 10$, and decaying probability $Np_t = \frac{1}{(1+t/100)^2}$

(c) Convergence in a network with 500 nodes, $m_u = 50$, and constant probability $Np_t = \frac{1}{2}$

Fig. 2. Convergence rate simulations. The neighbor set measurement $X_t$, for each node in the network, is shown as a function of the iteration number $t$.

that $\alpha_{m-1} \neq 0$. Assume $\lim_{T\to\infty} \prod_{t=0}^{T}(1-p_t) = c > 0$ (the limit exists, since it is a monotone bounded sequence), then

$$\lim_{T\to\infty} |\pi(T) - \mathbf{e}_m| = \left| \sum_{i=1}^{m-2} \alpha_i \xi^i \left( \lim_{T\to\infty} \prod_{t=0}^{T-1} \lambda_i(t) \right) + c\alpha_{m-1}\xi^{m-1} \right| \quad (9)$$

Since the eigenvectors are linearly independent, the RHS of (9) is non-zero. Thus, we have proved the theorem. ∎

*Corollary 4.1:* The graph generated by Algorithm 1 converges to a gradient topology with probability 1 if and only if

$$\lim_{T\to\infty} \sum_{t=0}^{T} p_t = \infty. \quad (10)$$

*Proof:* This follows from Theorem 4.1, and the relation

$$\lim_{T\to\infty} \prod_{t=0}^{T}(1-p_t) = 0 \Leftrightarrow \lim_{T\to\infty} \sum_{t=0}^{T} p_t = \infty$$

for $0 < p_t < 1$. ∎

### B. Convergence Rate Estimation

In this subsection, we investigate the convergence rate of $X_t$, with a constant sampling probability $p_t = p$. Define

$$T_i = \inf_t \{X_t = 1 \mid X_0 = i\}$$

as the first time when $X_t$ reaches 1, when starting with $X_0 = i$. Further, let $M_i = \mathbb{E}[T_i]$ denote the expected time of convergence. Clearly $M_1 = 0$, and for $i = 2, \ldots, m$ we have

$$M_i = 1 + \mathcal{P}\{X_{t+1} = i-1 \mid X_t = i\} \cdot M_{i-1}$$
$$+ \mathcal{P}\{X_{t+1} = i \mid X_t = i\} \cdot M_i$$
$$= 1 + (i-1)pM_{i-1} + (1-(i-1)p)M_i$$
$$\Rightarrow$$
$$M_i = \frac{1 + (i-1)pM_{i-1}}{(i-1)p} = \frac{1}{(i-1)p} + M_{i-1}$$

Continuing by induction yields

$$M_i = \frac{1}{p}\sum_{n=1}^{i-1} \frac{1}{n}.$$

The worst initial case is when $X_0 = m$, where the expected convergence time is

$$M_m = \frac{1}{p}\sum_{n=1}^{m-1} \frac{1}{n} \leq \frac{1+\ln(m-1)}{p}. \quad (11)$$

*Remark 4.1:* Notice that $M_m$ is the expected time for an individual node to converge, and not the expected time for all the nodes in the network to converge to a gradient topology.

## V. CONVERGENCE SIMULATION

In this section, we examine the convergence of Algorithm 1 with numerical examples. In all examples, the utility value set consists of ten distinct values, $\Lambda = \{1, \ldots, 10\}$. In the first two simulation (figure 2(a) and figure 2(b)) the number of nodes of each utility value is $m_u = 10$, and for the second simulation (figure 2(c)) $m_u = 50$. Thus, the total number of nodes in the network is $N = 100$ and $N = 500$ respectively.

The similar view $N_i^s(0)$ is initialized with $m_u$ nodes uniformly chosen among all nodes in the network. In the first and third simulation the sampling probability $p_t$ is held at a constant value of $\frac{1}{2N}$. Hence, for each node, and at each iteration of the algorithm, the random view is empty with probability $\frac{1}{2}$. Theorem 4.1 guarantees the convergence of the algorithm for these examples, which is also confirmed by the simulations. These two simulations should also be compared to the expected convergence rate given by equation (11), 566 and 4479 iterations respectively.

In the second simulation (figure 2(b)), we also analyze a decaying probability $p_t = \frac{1}{N}\frac{1}{(1+t/100)^2}$. Notice that $\sum_{t=0}^{\infty} Np_t < 101$, hence, by Corollary 4.1, there is a positive probability that the algorithm does not converge to a gradient topology. This is also confirmed by the simulation, in which the gradient topology is missing.

## VI. LIVE-STREAMING USING THE GRADIENT - EXPERIMENTS

Here, we evaluate the effect of sampling from the Gradient overlay compared to a random overlay when building a P2P live-streaming application called GLive. GLive is based on nodes cooperating to share a media stream supplied by a source node. GLive uses an approximate auction algorithm to match nodes that are willing and able to share the stream with one another. GLive extends our previous work on tree-based live-streaming, gradienTv [13] and Sepidar [9], to mesh-based live-streaming.

Nodes want to establish connections to other nodes that are as close as possible to the source. They bid for connections to the best neighbours using the upload bandwidth they contribute as money. Nodes share their bounded number of connections with nodes who bid the highest (contribute the most upload bandwidth). Auctions are continuous and restarted on failures or free-riding. The desired affect of our auction algorithm is that the source will upload to nodes who contribute the most upload bandwidth, who will, in turn, upload to nodes who contribute the next highest amount of bandwidth, and so on until the topology is fully constructed. More details on our approximate assignment algorithm can be found in [9].

One of the main problems with the lack of global information about nodes' upload bandwidths is that it affects the rate of convergence of auction algorithm. Nodes would ideally like to bid for connections to other nodes who they can afford to connect to, rather than win a connection to a better node and be later removed because a better bid was received. The traditional way to discover nodes (to bid on) is using a uniform random peer-sampling service [5]. Instead, we use the Gradient overlay to sample nodes, where a node's utility value is the upload bandwidth it contributes to the system. As such, the Gradient should provide other nodes with references to nodes who have well-matched upload bandwidths. In [9], we showed that using the Gradient overlay reduced the rate of parent switching for tree-based live-streaming by 20% compared to random peer sampling. Here, we show for GLive the effect of sampling neighbours using random peer sampling (GLive/Random) versus sampling from the Gradient overlay (GLive/Gradient).

We implemented GLive using Kompics' discrete event simulator that provides different bandwidth, latency and churn models. In our experimental setup, we set the streaming rate to $512Kbps$, which is divided into blocks of $16Kb$. Nodes start playing the media after buffering it for 5 seconds. The size of similar-view in GLive is 15 nodes. In the auction algorithm, nodes have 8 download connections. To model upload bandwidth, we assume that each upload connection has available bandwidth of $64Kbps$ and that the number of upload connections for nodes is set to $2i$, where $i$ is picked randomly from the range 1 to 10. This means that nodes have upload bandwidth between $128Kbps$ and $1.25Mbps$. As the average upload bandwidth of $704Kbps$ is not much higher than the streaming rate of $512Kbps$, nodes have to find good matches as parents in order for good streaming performance. The media source is a single node with 40 upload connections, providing five times the upload bandwidth of the stream rate. We assume 11 utility levels, such that nodes contributing the same amount of upload bandwidth are located at the same utility level. Latencies between nodes are modeled using a latency map based on the King data-set [14]. We assume the size of sliding window for downloading is 32 blocks, such that the first 16 blocks are considered as the in-order set and the next 16 blocks are the blocks in the rare set. A block is chosen for download from the in-order set with 90% probability, and from the rare set with 10% probability. In the experiments, we measure the following metrics:

1) *Playback continuity*: the percentage of blocks that a node received before their playback time. We consider the case where nodes have a playback continuity of greater than 99%;
2) *Playback latency*: the difference in seconds between the playback point of a node and the playback point at the media source.

We compare the playback continuity and playback latency of GLive/Gradient and GLive/Random in the following scenarios:

1) *Churn*: 500 nodes join the system following a Poisson distribution with an average inter-arrival time of 100 milliseconds, and then till the end of the simulations nodes join and fail continuously following the same distribution with an average inter-arrival time of 1000 milliseconds;
2) *Flash crowd*: first, 100 nodes join the system following a Poisson distribution with an average inter-arrival time of 100 milliseconds. Then, 1000 nodes join following the same distribution with a shortened average inter-arrival time of 10 milliseconds;
3) *Catastrophic failure*: 1000 nodes join the system following a Poisson distribution with an average inter-arrival time of 100 milliseconds. Then, 500 existing nodes fail following a Poisson distribution with an average inter-arrival time 10 milliseconds;

Figures 3 shows the percentage of the nodes that have playback continuity of at least 99%. We see that all the nodes in GLive/Gradient receive at least 99% of all the blocks very quickly in all scenarios, while it takes slightly more time for GLive/Random. That is because nodes in GLive/Random randomly sample nodes to run the auction algorithm against, while GLive/Gradient runs the auction algorithm against nodes that contribute similar amounts of upload bandwidth. Random sampling takes longer time to find good matches for delivering the stream. One point to note is that the 5 seconds of buffering cause the spike in playback continuity at the start, which then drops off as nodes are joining the system. To summarize, using the Gradient overlay instead of random sampling produces better performance when the system is undergoing large changes - such as large numbers of nodes joining, failing over a short period of time. Figure

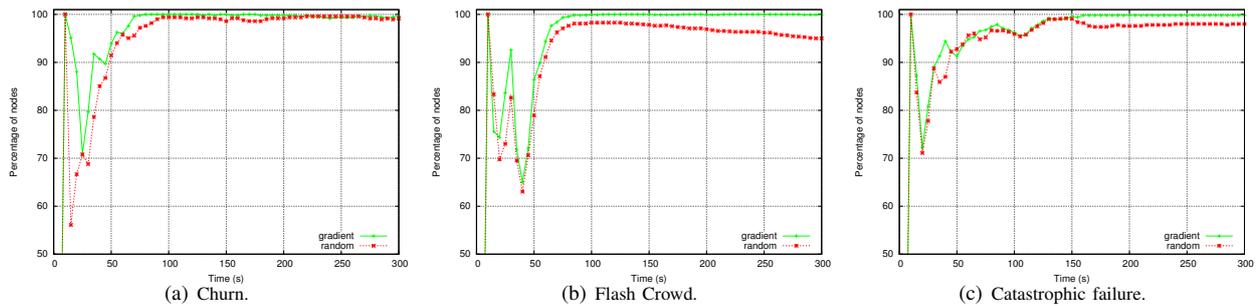

Fig. 3. Playback continuity of the systems in different scenarios.

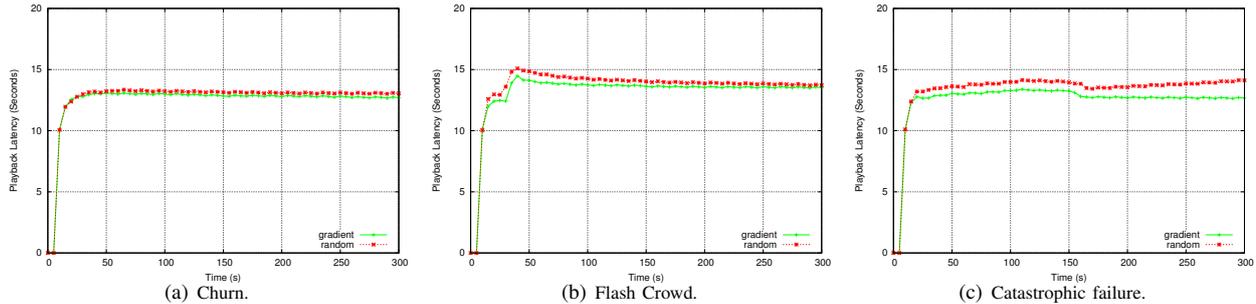

Fig. 4. Playback latency of the Gradient versus Random sampling in different scenarios.

4 shows the playback latency of the systems in the different scenarios. As we can see, although there is only a small difference between the systems, although, GLive/Gradient consistently maintains relatively shorter playback latency than GLive/Random for all experiments. The playback latency includes both the 5 seconds buffering time and the time required to pull the blocks over the live-streaming overlay constructed using the auction algorithm.

## VII. CONCLUSIONS

In this paper, we introduced the topology convergence problem for the gossip-generated Gradient overlay network. We showed the necessary and sufficient conditions for convergence to a complete gradient structure We characterized the convergence time and provided bounds on the worst-case convergence time. Our experiments show the potential advantages of topologies built using preference functions. We showed how nodes can use implicit information captured in the Gradient topology to more efficiently find suitable neighbours compared to random sampling. As such, our work on proving convergence properties of the Gradient topology should have significance for other future information-carrying topologies. In future work, we will examine modifications to the topology construction algorithm that improve convergence time, as well as further applications of the topology in building P2P applications.